\documentclass[twocolumn,showpacs,preprintnumbers,amsmath,amssymb]{revtex4}

\def\pra{{Phys.~Rev.~A}}

\usepackage{isomath}

\usepackage{upgreek} 

\usepackage{epsfig, amsmath,amsfonts, amssymb,graphicx,color}

\usepackage{mathtools}
\usepackage{dcolumn}
\usepackage{bm}
\usepackage{rotating}
\usepackage{amssymb}
\usepackage[latin1]{inputenc}

\begin{document}

\title{Laser cooling of molecular anions}
\author{Pauline Yzombard$^{1}$, Mehdi Hamamda$^1$, Sebastian Gerber$^2$, Michael Doser$^2$  and Daniel Comparat$^1$}
\affiliation{$^1$ Laboratoire Aim\'{e} Cotton, CNRS, Universit\'{e} Paris-Sud, ENS Cachan, B\^{a}t. 505, 91405 Orsay, France \\ $^2$CERN, European Laboratory for Particle Physics, 1211 Geneva, Switzerland }

\date{\today}

\begin{abstract}
We propose a scheme for laser cooling of negatively charged molecules.
We briefly summarise the requirements  for such laser cooling and we identify a number of potential candidates.
A detailed computation study with C$_2^-$, the most studied molecular anion, is carried out. Simulations of 3D laser cooling in a gas phase show that this molecule could be cooled down to  below  1 mK in only a few tens of milliseconds, using standard lasers.
Sisyphus cooling, where no photo-detachment process is present,  as well as Doppler laser cooling of trapped C$_2^-$, are also simulated.
This cooling scheme has an impact on the study of cold molecules, molecular anions, charged particle sources and antimatter physics.
\end{abstract}

\maketitle


Molecular anions play a central role in a wide range of fields: from the chemistry of highly correlated systems
\cite{Bell2009,2009RPPh...72h6401D,Review_Ye2009,2012_JinYe_Introduction} to atmospheric science to the study of the interstellar-medium \cite{2006CPL...431..179L,simons2008molecular,2006ApJ...652L.141M,2014IAUS..297..344F,2011ARPC...62..107S}. However,
it is currently very difficult to investigate negative ions in a controlled manner at the ultracold temperatures relevant for the processes in which they are involved. Indeed, at best, temperatures of a few kelvins have been achieved using supersonic beam expansion methods or trapped particles followed by electron cooling, buffer gas cooling or resistive cooling  \cite{gerlich1995ion,PhysRevLett.82.3198,2012PhRvA..86d3438D,2013ApJ...776...25K,2014Icar..227..123B}.
The ability to cool molecular anions to sub-K temperatures would finally allow investigation of their chemical and physical properties at   energies appropriate to their interactions.
Furthermore, anionic molecules at mK temperatures can also play an important role in (anti)atomic physics, where copious production of sub-K antihydrogen atoms currently represents the dominant challenge in the field.
Sympathetic cooling of antiprotons via laser-cooled atomic negative ions that are simultaneously confined in the same ion
trap has been proposed as a method to obtain sub-K antiprotons \cite{2006NJPh....8...45K}. As an alternative to this yet-to-be-realized procedure, ultra-cold molecular anions could replace atomic anions in this scheme, and would thus facilitate the formation of ultra-cold antihydrogen atoms.
More generally, cooling even a single anion species would be the missing tool to cool any other negatively charged particles (electrons, antiprotons, anions) via sympathetic cooling.

In this letter we present a realistic scheme for laser cooling of molecular anions to mK temperatures.
Laser cooling of molecules has been achieved only for a very few neutral diatomic molecules (SrF, YO, CaF) \cite{shuman2010laser,2013PhRvL.110n3001H,2013arXiv1308.0421Z}. Furthermore, even if well established for neutral atoms and atomic cations, laser cooling techniques have so far never been applied to anions \cite{PhysRevA.89.043430}.  This is because in atomic anions, the excess electron is only weakly bound by quantum-mechanical correlation effects. As a result, only a few atomic anions are known to exhibit electric-dipole transitions between bound states: Os$^-$, La$^-$ and Ce$^-$ \cite{2010PhRvA..82a4501P,2011PhRvA..84c2514W,2014PhRvL.113f3001W}.

For molecules the situation is quite different because their electric dipole can bind an extra electron, and even di-anions have been found to exist \cite{schauer1990production}. For instance, polar molecules with a dipole exceeding 2.5 Debye exhibit dipole-bound states. Highly dipolar molecules such as LiH$^-$, NaF$^-$ or MgO$^-$ possess several such dipole-bound states \cite{1984PhRvL..52.2221L,1997CPL...276...13G}, and valence anionic states exist as well.
For simplicity, in this letter we only focus on diatomic molecules, even if the Sisyphus laser cooling techniques proposed here can also be applied to trapped polyatomic molecules \cite{2012Natur.491..570Z}.
 In the Supplementary material, Table \ref{tab:mol} \cite{Supplement}, we present a review of most of the experimental as well as theoretical studies of diatomic anions, with useful references, if further studies are required.

The first excited state (and sometimes even the ground state) of many molecular anions lie above
their neutralization threshold, such as in the case of H$_2^-$, CO$^-$, NO$^-$, N$_2^-$, CN$^-$ or most of systems with 3, 4 or 11 outermost electrons.
For this reason, these anions are not stable against auto-detachment processes and exhibit pure ro-vibrational transitions with $\sim$ $100 \, $ms lifetimes. Even if such long-lived states can still be of interest for Sisyphus cooling, for narrow-line cooling or for Doppler laser cooling in traps  \cite{2012Natur.491..570Z,2014PhRvA..89d3410C}, pure electronic transitions are preferred for rapid laser cooling.  Transitions of $\sim$ $100\, $ns lifetime can be found between well-separated electronic states (typically B$\leftrightarrow$X states), whereas
 transitions in the infrared region between electronic states (typically A$\leftrightarrow$X  states) have longer lifetimes of $\sim$ $100 \, \upmu$s.
Some challenges in laser-cooling molecular ions have been described in \cite{2011NJPh...13f3023N}.
For direct laser cooling of neutral diatomic molecules, a key ingredient is a good branching ratio, i.e. Franck-Condon factor (FC factor), between vibrational levels
(SrF, YO, CaF all have more than a 98\% branching ratio on the $A^2\Uppi_{1/2} (v'=0, J'= 1/2)  \leftarrow X^2 \Upsigma (v''=0, N''=1)$ transition).
Even with these considerations, the choice of the most suitable candidate is not obvious and is a compromise between using fast electronic transitions and choosing extremely good FC factors.
Indeed, molecules with good FC factors can be found for weak dipole-valence bound transitions or for \textcolor{black}{anionic} molecules with 6 or 12 outermost electrons but with forbidden dipole transitions.
FC factors greater than 70\%  can be found in systems with 8, 14 or higher numbers of outermost electrons \textcolor{black}{or in molecules which include Li or Al atoms \cite{1993JChPh..99.8793B,1999JChPh.110.2928G}.  }
Unfortunately the corresponding transitions are often in the deep infrared region.
The best compromise seems  to be the systems with 9 outermost electrons  having  X$ (\upsigma^1 \uppi^4) ^2 \Upsigma_{(g)}^+$, A$(\upsigma^2 \uppi^3) ^2 \Uppi_{(u)}$,   B$(\upsigma^2 \uppi^4) ^2 \Upsigma_{(u)}^+$  levels. A list of such systems is given in Table \ref{tab:mol}.

Clearly the most studied system  is  C$_2^-$, with a perfectly known spectrum. C$_2^-$ exhibits a B$^2\Upsigma (v'=0)  \leftrightarrow $X$^2 \Upsigma (v''=0)$  system and a  A$^2\Uppi_{1/2} (v'=0)  \leftrightarrow$ X$^2 \Upsigma (v''=0)$ system with 72\% and 96 \% branching ratio, respectively. Besides, this anion has the notable advantage of not presenting any hyperfine structure.
As a potential further benefit of this system, we mention that
through laser photo-detachment of cold C$_2^-$, we could produce cold C$_2$ molecules, important in combustion physics and astrophysics. Even if more studies on laser cooling are needed, C$_2$ looks like a suitable candidate to be further laser cooled near 240 nm on the 0-0 Mulliken band  (d $^1\Upsigma_{\rm u}^+ \leftarrow $X$^1\Upsigma_{\rm g}^+$) which has an extremely favorable branching ratio of 99.7 \% \cite{sorkhabi1998franck}.

	\begin{table}[htbp]
					\centering
					\footnotesize
	\begin{tabular}{|l|l|l|l|l|}
	\hline\noalign{\smallskip}
 group & example  \\
	\hline\noalign{\smallskip}
		\hline\noalign{\smallskip}
 I-VII & LiF$^-$ LiCl$^-$ NaF$^-$ NaCl$^-$ MnH$^-$(X$^6\Updelta$)    \\
 II-VI &   ZnO$^-$
BeO$^-$
 MgO$^-$
ZnF$^-$   \\
 III-V & BP$^-$ AlN$^-$ AlP$^-$  AlAs$^-$
 GaP$^-$ InP$^-$  GaAs$^-$
 BN$^-$ GaN$^-$  \\
 IV-IV &  C$_2^-$
	Si$_2^-$ CSi$^-$(B unstable) Sn$_2^-$  Pb$_2^-$ SnPb$^-$(X$^2 \Uppi$)  \\
				\hline\noalign{\smallskip}
\end{tabular}
				\caption{Diatomic anions with 9 outermost electrons, listed in terms of group of atoms. A more complete table with all diatomic molecules, and references, is given in the Supplementary material. They are isoelectronic to the neutral molecules CN and SrF with the same  X$^2 \Upsigma^+$ A$^2 \Uppi$   B$^2 \Upsigma^+$ electronic structure  with sometimes stable quartet excited states. }
				\label{tab:mol}
			\end{table}

We will therefore concentrate on C$_2^-$ as a benchmark to study laser cooling of anionic molecules. Note however that several other molecules, such as BN$^-$  or AlN$^-$, may be used as well. They offer very similar structure probabilities with better FC factors (higher than 98\%) but
with a    B$^2 \Upsigma \rightarrow $A$^2 \Uppi $ decay channel that is absent in the homonuclear case of C$_2^-$.  Contrary to C$_2^-$,  heteronuclear molecules present a closed rotational level scheme.  However, further spectroscopic studies are clearly required for such systems, as well as for other
promising ones, such as metal-oxyde systems (FeO$^-$, NiO$^-$) or hydride ones (CoH$^-$).

\begin{figure}[hb]
\mbox{\resizebox{0.45\textwidth}{!}{\includegraphics{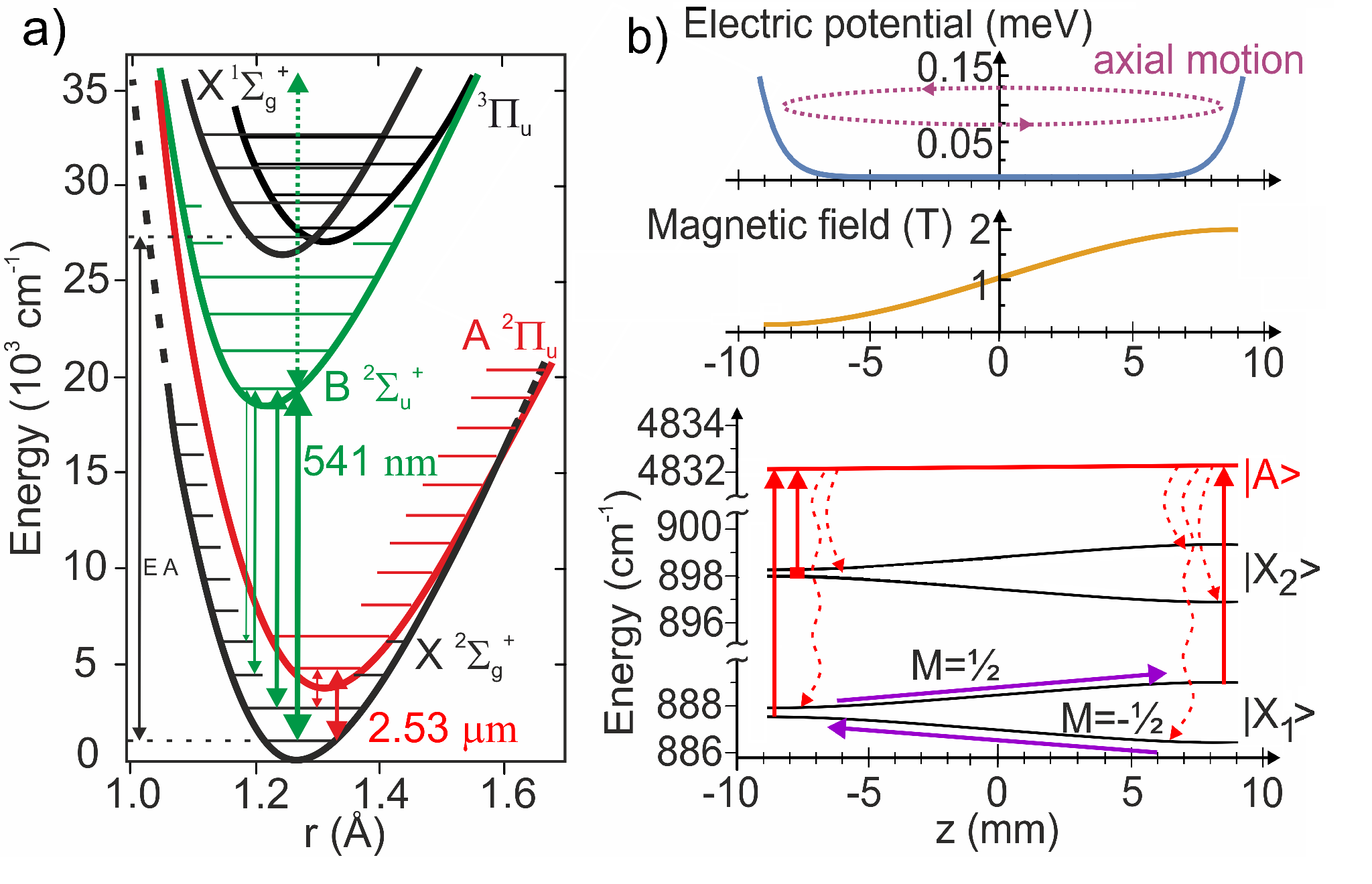}}}
\begin{center}
		\caption{(Colour online). 	C$_2^-$ level structure and laser cooling transitions. (a): Electronic and vibrational levels including the photo-detachment  threshold of neutral C$_2(^1\Upsigma_{\rm g}^+, ^3\Uppi_{\rm u})$, from \cite{ervin1991photoelectron}. The widths of the arrows are proportional to the transition strengths given by the FC.
The green dotted line indicates that photo-detachement can occur from the B state due to a second photon absorption.
 (b): Detail of the Sisyphus cooling principle with a zoom on the X $\leftrightarrow$ A energies: \textcolor{black}{ $\mid$X$_1$$\rangle$=X($v''=0,N''=0$) and $\mid$X$_2$$\rangle$=X($v''=0,N''=2$),}  $\mid$A$\rangle$=A($v'=0,N'=1,J'=1/2$). A Penning-like trap is presented in the upper part. A magnetic gradient field creates the Sisyphus potential hill. The corresponding Zeeman effect on the C$_2^{-}$ internal level states is plotted in the lower part. Laser excitations and spontaneous decays are respectively illustrated with red solid line and red dashed line.  }
\label{C2anion}
\end{center}
\end{figure}

In order to study laser cooling of C$_2^-$ we have performed three types of simulations. The first one is "standard" laser cooling in the gas phase (thus with no strong external fields present); the second one is laser cooling in a Paul trap; and the last one is
Sisyphus cooling of trapped ions in a Penning trap. The simulations are performed with the
C++  code described in
\cite{2014PhRvA..89d3410C}, which now also includes full N-body space charge effects \cite{dehnen2011n}; the Lorentz force is solved by using Boris-Buneman integration algorithms
\cite{zwicknagel2008simulations,toggweiler2014novel}. Briefly, a Kinetic Monte Carlo
algorithm gives the exact time of events (absorption or emission of light) when solving the rate equations to study laser
excitation of the molecules under the effect of Coulomb, light scattering, dipolar, Stark and Zeeman forces.
The C$_2^-$ energy levels and required laser transitions are shown in Fig. \ref{C2anion}(a).

The lifetime of the B-state is 75 ns \cite{1984_lifetime_C2_anion,pot_C2_anion_2006} and wavelengths from its first vibrational level to the X state vibrational levels are 541, 598, 667, 753, 863, 1007, 1206 nm with a corresponding transition strength  probability of 72, 23, 5, 0.8, 0.1, $2\times 10^{-4}$, $3\times 10^{-5}$, $4\times 10^{-6}$ (percentage of the FC), from \cite{jones1980photodetachment,shan2003study}.
The	lifetime of the A-state is 50 $\upmu$s with wavelengths of 2.53 $\upmu$m  (branching ratio of 96\%) and 4.57 $\upmu$m (branching ratio of 4\%).
In order to close the rotational transition cycle, two lasers are required for each of these vibrational transitions \textcolor{black}{to couple X($N''=0,\,2)$ to A($N'=1,J'=1/2$)}, see Fig. \ref{C2anion}(b).

For our simulations, the temperature along one axis is calculated from the deviation of 50\% of the central velocities' histogram. The so-called 3D-temperature ($T_{3D}$) is the quadratic mean of the three one-dimensional temperatures (x,y,z).

	For our first simulation, we study a possible experiment based on deceleration of an anionic beam.
	In contrast to neutral cold molecules which are
difficult to produce at low velocity, and in spite of the development of techniques such as velocity filtering, buffer gas cooling or decelerators (see list in \cite{2014PhRvA..89d3410C}), an anionic beam can be brought to a standstill very easily by an electric field \cite{RSIGonzalez2013}.
Indeed, a typical beam of C$_2^-$ has a current of 1 nA and is emitted at 1000 eV with an energy dispersion of 1 eV \cite{PhysRevA.28.1429}.  A 1000 volt potential box can thus decelerate such a beam. Furthermore,
due to the energy dispersion of the beam, the stability of this voltage power supply is not critical.
Typically 1/100000 of the anions (i.e. 0.01 pA current) will be decelerated inside the box to below an energy of 0.01 meV. This corresponds roughly to 0.1 K, which is within the capture range of molasses cooling.

Thus, we propose a very simple  deceleration scheme using a grounded vacuum chamber through which the 1000 eV beam  propagates; this is followed by a chamber at 900 V with a 3 mm radius hole which focuses the beam (1 cm downstream) to a smaller hole of R $ \sim 0.3 $ mm radius of an innermost chamber at 1000 V. The chambers could be made of  transparent conducting film (such as Indium Tin Oxide) or as grids to allow laser cooling.  In this final chamber the electric field decreases as $\sim$ $ 0.2  (z/R)^{-3}$ along the propagation axis z \cite{1999NIMPA.427..363R} such that at 1 cm the Lorentz force
 becomes negligible compared to the laser cooling force that is typically
 $\hbar k \Gamma/10 \sim 10^{-20} $N, corresponding to an acceleration of $10^5 $m $\cdot$ s$^{-2}$.

\begin{figure}[t]
\mbox{\resizebox{0.45\textwidth}{!}{\includegraphics{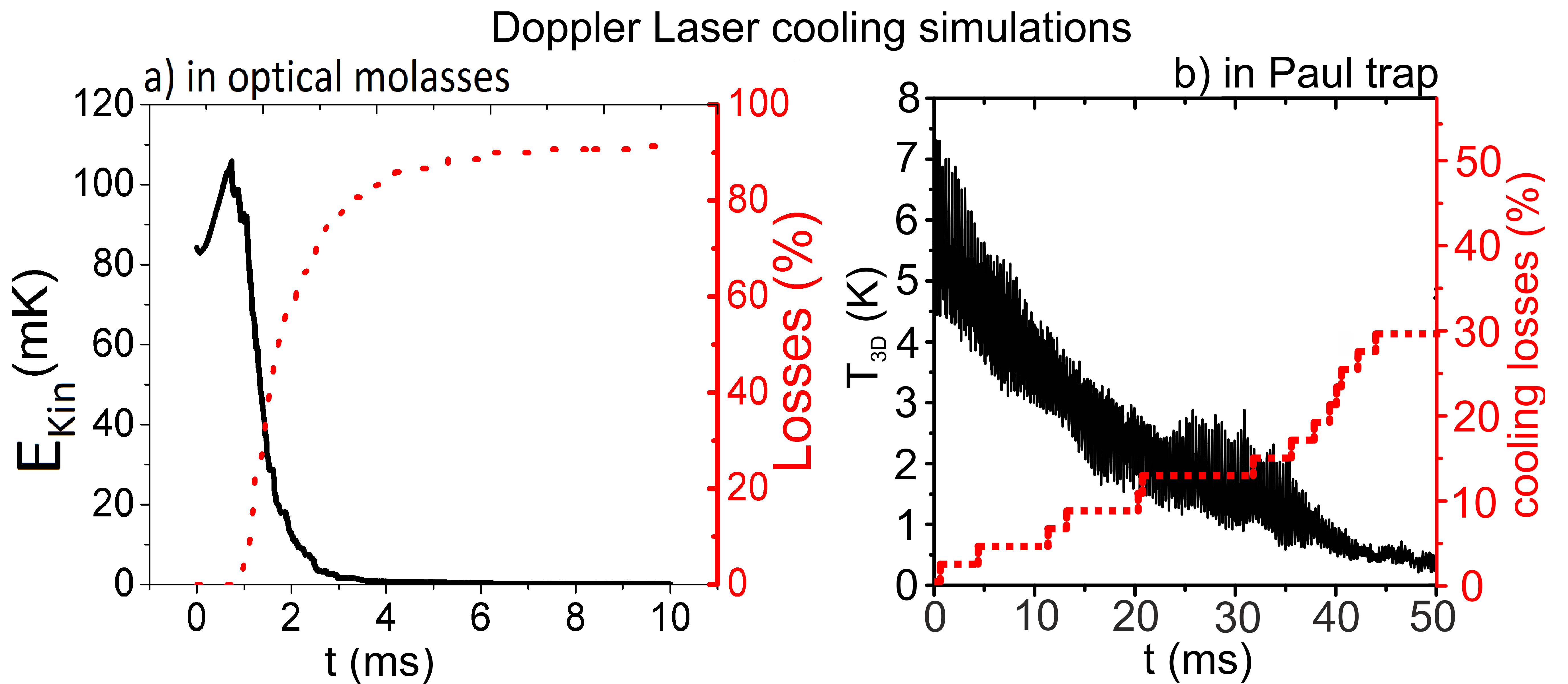}}}
\begin{center}
		\caption{(Colour online). Doppler laser cooling results, cycled on the X$\leftrightarrow$B transition. (a):  Deceleration and 3D molasses cooling of C$_2^-$. Result of the 3D laser cooling simulation of an initial bunch of 150 anions at 80
mK: The red dashed curve gives the population losses, corresponding to molecules which escape from the laser interaction area. Black curve: kinetic energy evolution (in temperature unit) of the remaining particles.
		(b): In black, evolution of the temperature and number of C$_2^-$ initially at 5K in a Paul trap. The red dashed curve gives the population losses, through photo-ionisation or decays into high vibrational ground states ($v>4$).}
\label{fig:doppler_cooling}
\end{center}
\end{figure}

We simulate such a 3D molasses cooling with several CW lasers, all with 0.5 W power and 2 cm waists and wavelengths 10 MHz red detuned from the transitions (Fig. \ref{C2anion}). We consider here a cooling on the \textcolor{black}{ X$(v''=0,N''=2) \leftrightarrow $B$(v'=0, N'=0)$ }transition.
For the 3D cooling we thus have 6 lasers on the \textcolor{black}{X$(v''=0,N''=2) \leftrightarrow $ B$(v'=0, N'=0)$ }transition, 2 repumping counter-propagating lasers on the \textcolor{black}{X$(v''=0, N''=0) \leftrightarrow $B$(v'=0,N'=0)$}, plus extra repumping lasers: 4 lasers for each higher vibrational level of the ground state X up to $v''=4$ (these lasers are counter-propagating along $+/-$ Z).
We start with 150 particles at 80 mK distributed following  a spherical gaussian distribution ($\sigma = 4$ mm).
The result is indicated in Fig. \ref{fig:doppler_cooling}(a).  The first increase at 1 ms is due to the conversion of the Coulomb potential energy into kinetic energy. Then, a fast cooling down to 1 mK is observed, although many molecules escape from the laser waist interaction area.
We find that the final temperature and losses depend on the initial density (initial gaussian radius) because of space charge effects that counter the cooling.
Note that we do not attempt trapping but only molasses-cool here. Indeed, realizing a trap (MOT)  would require a magnetic-field gradient producing a Lorentz force  stronger than the laser trapping one, especially because the Zeeman effect of \textcolor{black}{B$(v'=0, N'=0)$ }is weak (quadratic behavior).
 To be feasible, this cooling method requires pre-cooling down to few hundreds of mK in order to avoid having too fast molecules.

In the second simulation we study particles in a Paul trap. For simplicity, we only consider motion
 in an harmonic pseudopotential well $V({\bf r}) = 1/2\ m\ \omega_r (x^2 + y^2) + 1/2\ m\ \omega_z^2$
\cite{PhysRevA.76.012719}. We chose 5 K as a typical starting temperature of the beam. This would correspond to having carried out a first cooling step, using e.g. Helium as a buffer gas.
We simulate Doppler cooling on the  X  $\leftrightarrow$ B transitions, with a 600 MHz red detuning, and a spectral laser bandwidth of 35 MHz. As previously, we consider repumping lasers from the different vibrational levels of the ground state X up to X($v''=4$), as shown in Fig. \ref{C2anion}(a). But in this case repumping lasers are only along +Z, since particles are trapped.
Results are given in Fig. \ref{fig:doppler_cooling}(b) where cooling down to 60 mK is achieved within 50 ms.
As the photo-detachment cross-section of C$_2^-$ for the B-state is unknown, we use the photo-detachement
cross section of C$_2$,
$\sigma \sim 10^{-17}$ cm$^2$
\cite{2013ApJ...776...25K} as typical value.
 The photo-detachement rate for the B-state is $ I \sigma /h \nu = 4.3 $ s$^{-1}$ for $I = 0.16 $ W/cm$^2$. Within 50 ms of cooling we  loose only 3\% of the molecules by photo-detachement.
To these photo-detachement losses, we have to add 25\% of decays to higher vibrational levels of the X-state. The losses' evolution over time is shown in Fig. \ref{fig:doppler_cooling}(b).
Here, we would like to emphasize that for both molasses cooling and Paul trap simulations, we use a Doppler laser cooling process, with the same lasers. The Paul trap can thus serve as a first cooling and can be turned off for further cooling using the molasses cooling for low density clouds.
Photo-detachement and decays in higher vibrational X-state levels are thus similar. For the simulations of molasses cooling however, losses due to motion of molecules out of the laser's area are much more important than photo-detachment or decay losses.

\begin{figure}
\mbox{\centering\resizebox{0.45\textwidth}{!}{\includegraphics{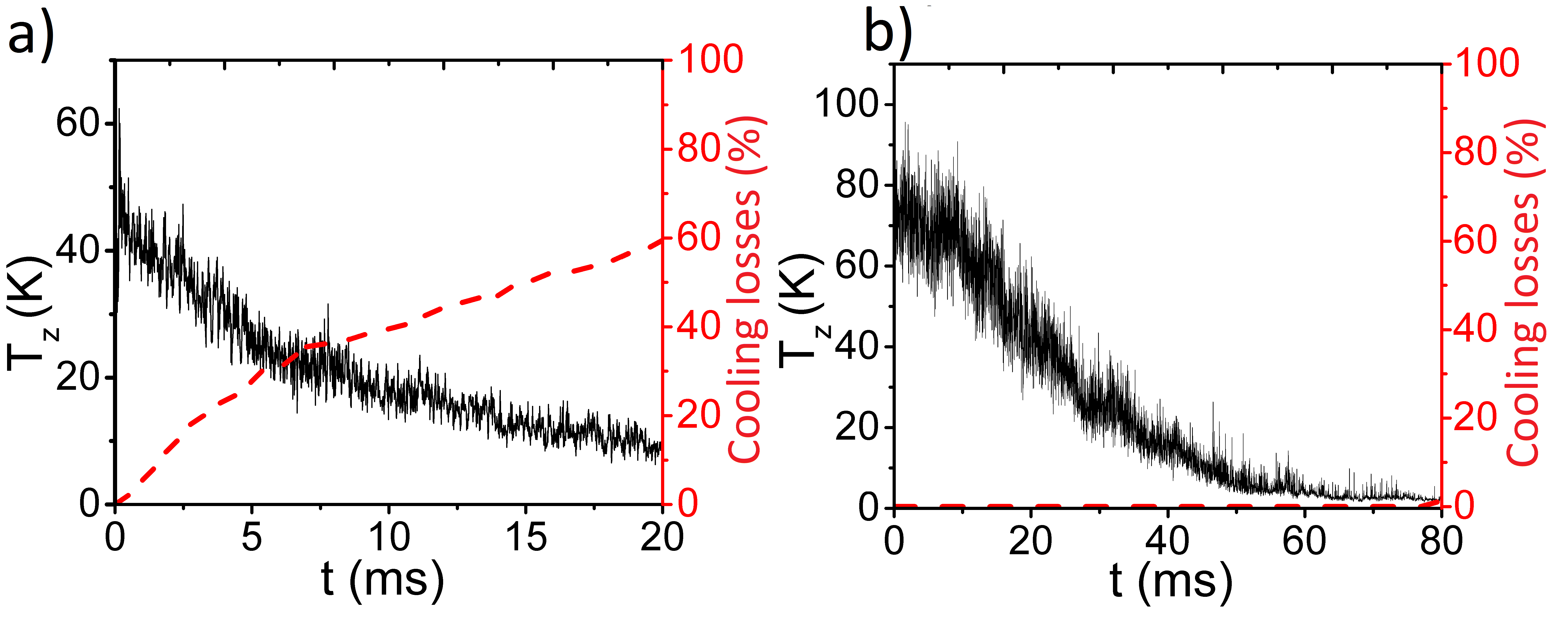}}}
		\caption{(Colour online).  Sisyphus cooling curves corresponding to the evolution of the axial temperature and number of C$_2^-$ initially at 60 K and 100 K in an inhomogenous Penning trap and cooled on the X$\leftrightarrow$A transition, with (a) or without (b) repumping the X($v''=1$)losses.}
\label{fig:sisyphus}
\end{figure}

A solution to avoid the losses of C$_2^-$ through photo-detachment or decays into high vibrational levels of the ground-state is to cool through the A-state.
This cooling has very similar characteristics (linewidth, wavelength) to those of La$^-$ \cite{2014PhRvL.113f3001W} but with a lighter particle and no photo-detachment in the case of C$_2^-$.

 Our third simulation thus concentrates on cooling and trapping  in a $ \sim\, $1.5 cm long Penning-like trap using Sisyphus-type cooling \cite{2014PhRvA..89d3410C}.
The principle is illustrated in Fig. \ref{C2anion}(b): due to the axial motion induced by the electric trapping potential (300 $\upmu$s oscillation time in the simulation) in a Penning-like trap, particles move between the high (2 T) and low (0.2 T) magnetic fields at both ends of their axial range. A given particle is initially in the \textcolor{black}{X($v''=0, N''=0, M''=1/2$)} state; then, after an absorption followed by spontaneous emission, it decays towards the \textcolor{black}{X($v''=0, N''=0, M''=-1/2$)} state. More than 1 K is removed for each closed absorption-emission cycle, as the molecule continuously climbs the magnetic potential hill.

In the last 3D simulation we focus on  the axial temperature, set initially at 100 K for a cloud of 200 anions.
For our low density plasma there is  no coupling between radial and axial motions, but the radial shape of the plasma reflects the inhomogeneous magnetic field.
Two cooling lasers at\textcolor{black}{ $2.5351$ $\upmu$m and $2.5358$  $\upmu$m }with $100$ MHz spectral bandwidth each are detuned to be resonant at $0.2$ T and $2$ T, respectively, along the \textcolor{black}{ X($v''=0,N''=0, M''=-1/2$)$_{0.2T}$ } $\leftrightarrow$ A($v'=0, N'=1, J'=1/2$)$_{0.2T}$ and \textcolor{black}{X($v''=0,N''=0,M''=1/2$)$_{2T}$ } $\leftrightarrow$ A($v'=0, N'=1, J'=1/2$)$_{2T}$ transitions. The considered laser power is 0.05 W for a \textcolor{black}{1 mm} waist.
To avoid relying on too many lasers, we repump the losses in the \textcolor{black}{X($v''=0,N''=2$)} states with only one single broadband (6000 MHz, 0.05 W) laser which addresses all the Zeeman-split sub-levels, resonant at 0.2 T.

	Results are given in Fig.\ref{fig:sisyphus}(a). In tens of ms, the axial temperature is cooled down from 60 K to few K. The lost population mainly goes to X($v''=1$).  As for all simulations, we load the trap using an initial (non thermalized) Gaussian velocity distribution. The evolution of this non equilibrium system leads to the high frequency velocity (and thus instantanous temperature) fluctuations in both Fig.\ref{fig:doppler_cooling}(b) and Fig.\ref{fig:sisyphus}.
We also present an alternative simulation, of which results are given in Fig. \ref{fig:sisyphus}(b). Here, decays in the vibrational X($v''=1$) states are repumped: two lasers, at $4.574 \ \upmu$m and $4.593\ \upmu$m, address the \textcolor{black}{ X($v''=1,N''=0$) $\leftrightarrow$ A($v'=0,N'=1,J'=1/2$) and X($v''=1,N''=2$) $\leftrightarrow$ A($v'=0,N'=1,J'=1/2$).} Both are resonant at $0.2$ T, with \textcolor{black}{both spectral bandwidth of $6$ GHz and power of 0.1 W.}

This simulation requires 2 more lasers but has the  advantage of cooling a greater part of the molecules (less than 0.5\% of anions fall in the X($v''=2$) levels, \textcolor{black}{within 80 ms, in comparison to the 60\% of losses within 20 ms, for the first case without repumpers on X($v''=1$)).}

In conclusion, we have presented several possible deceleration and laser cooling schemes for anionic molecules, either in free space or trapped, by using Doppler or Sisyphus cooling, circumventing the problem of photodetachment.
Working with traps can open many possibilities due to the long trapping times : it could enable restricting studies to only ro-vibrational transitions, or working with electronic transitions that have long spontaneous emission times. Furthermore, laser cooling of molecular anions followed by laser photo-detachment could be used as a source for cold neutral molecules, or as an ideal source for electron bunches, since no Coulomb force due to ions will be present and ideal uniform elliptical density shapes could be realized \cite{PhysRevLett.93.094802,1999RvMP...71...87D}.

Acknowlegments:
We are indebted to A. Kellerbauer, V. Kokoouline, O. Dulieu, C. Drag, M. Raoult and H. Lignier for useful discussions.
The research leading to these results has received funding from the European
Research Council under the grant agreement n.~277762 COLDNANO.


\end{document}